# Does Python Smell Like Java?
## Tool Support for Design Defect Discovery in Python

Nicole Vavrová[a] and Vadim Zaytsev[a,b]

a    Universiteit van Amsterdam, The Netherlands
b    Raincode Labs, Belgium

**Abstract**
   The *context* of this work is specification, detection and ultimately removal of harmful patterns in source code that are associated with defects in design and implementation of software. In particular, we investigate five code smells and four antipatterns previously defined in literature. Our *inquiry* is about detecting those nine design defects in source code written in Python programming language, which is substantially different from all prior research, most of which concerns Java and other C-like languages. Our *approach* was that of software engineers: we have processed existing research literature on the topic, extracted both the abstract definitions of defects and their concrete implementation specifications, programmed them all in a tool and let it loose on a huge test set obtained from open source code from thousands of GitHub projects. When it comes to *knowledge*, we have found that more than twice as many methods in Python can be considered too long (statistically extremely longer than their neighbours within the same project) than in Java, but long parameter lists are seven times less likely to be found in Python code than in Java code. We have also found that Functional Decomposition, the way it was defined for Java, is not found in Python code at all, and Spaghetti Code and God Classes are extremely rare there as well. The *grounding* and the confidence in these results comes from the fact that we have performed our experiments on 32,058,823 lines of Python code, which is by far the largest test set for a freely available Python parser. We have also designed the experiment in such a way that it aligned with prior research on design defect detection in Java in order to ease the comparison if we treat our own actions as a partial replication. Thus, the *importance* of the work is both in the unique open Python grammar of highest quality, applied to millions of lines of code, and in the design defect detection tool which works on something else than Java.

# The Art, Science, and Engineering of Programming

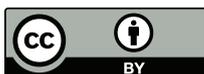

**Does Python Smell Like Java?**

## 1 Introduction

It is a well-known and widely accepted fact of programmers' lives that maintenance activities take up significant fraction of software development time. Software design defects — bad smells, antipatterns, bug patterns, pitfalls, convention violations, simply bad practices — are fragments of source code that have a noticeable negative impact on software maintenance. Most of the research conducted in the field of design defect detection is focused on Java source code. Java is certainly a popular language as it has held the first place in the well-known TIOBE programming community index [60] for a long time. However, other languages exist and matter, and in this paper we go down to number 5 in that list — namely, Python. It is significantly different from all languages above it (Java, C, C++ and C#) in many aspects, especially in those that influence detection of design defects. We wanted to investigate whether there are any notable differences from Java-based findings.

To detect design defects in Python, we have developed a tool called Design Defect Detector [61]. This tool is compatible with any Python version used as of 2017, from Python 2.5 up to Python 3.6. To achieve this, we have engineered a Python grammar in ANTLR4 [50, 51], converging several available grammars from various sources. To answer our questions (see section 1.2), this tool was used on a data set of 4,121 GitHub repositories, consisting of 32,058,823 lines of Python code. As a result, we have found that 8 out of 9 design defects we chose for our study, were detectable in Python. We also found that code smells were more common in Python than antipatterns. Comparing our results to DECOR [41], we have found that the density of detected design defects was slightly lower in Python code than in Java code.

There exist initiatives similar to ours, such as Pylint [59], which detects deviations from coding standards and looks for code smells. However, to the best of our knowledge, there is currently no scientific study of detecting design defects in Python.

**1.1 Motivation**

Most software systems go through constant evolution derived from the need to keep up with the changing requirements. Since design flaws are known to have a large negative impact on maintenance [7], this introduces a requirement for quality design of software. There exist multiple well-known, industry established design practices, such as design patterns [17, 19], design heuristics [53], best practices [6, 35, 38, 39], language design patterns [24, 49, 65, 66], implementation patterns [5, 50], all existing to aid the process of creating a good design.

In our work, we focus on the counterparts of good design practices, calling them *design defects*, since the earlier term "pitfall" [63] is too vague. Moha et al. defined design defect as "the embodiment of bad design practices in the source code of programs" [42]. This means that design defects span across all levels of software design. They include low level issues such as code smells [18], but also architectural issues, such as antipatterns [7]. Somewhere between them there are also design defects that violate coding conventions [20], but they are even less researched, understood and classified yet.

Detection of the different types of design practices aids in determination of the quality of software design. The work that has been done in this field is described in further detail in section 1.3. Design defect detection is challenging for numerous reasons. It is difficult to do manually due to the analysed systems often being very large. Most of the used techniques therefore employ various levels of automation. They can span across different subsystems, so they cannot be detected locally [11]. However, the largest difficulty with detecting design defects is that they are defined in a very loose manner. Unlike design patterns, which have concrete descriptions and precise UML definitions, design defects descriptions are mostly textual, written in a natural language and often contain phrases open to interpretation, such as "class with a large number of attributes, operations, or both" [7]. Such definitions are not only ambiguous, but also very context dependent — a normal sized class in one project or programming language can be considered a large class in another project or language. Brown et al. actually define an antipattern as a "pattern in an inappropriate context" [7].

Most of the design defect detection studies that currently exist, have focused on Java in their experiments [14, 15, 16, 25, 26, 40, 41, 42, 43, 46, 47, 48, 57, etc]. There are a couple of approaches that claim language independence, such as the ones by Llano and Pooley [33] and Cortellessa [12] — however, neither of them seems to have gained a lot of popularity.

We consider the context of a programming language to be a significant one. Java is a relatively statically typed and relatively strictly object-oriented language known for its verbosity. Python can occasionally be used for solving the same tasks, but it is much more concise, significantly more volatile when it comes to types, and it allows freeform switching between the scripting, imperative, functional and object-orientation paradigms. Thus, the ambition behind this paper is to partially fill the gap in design defect detection research in programming languages other than Java.

**1.2 Goals and Expectations**

The main goal of this project is to create a tool, Design Defect Detector [61], which automatically detects design defects in Python, and use it in an experiment. For the development of this tool, there are numerous considerations to be taken into account.

Detecting design defects requires a top-down approach, meaning the design defects to be detected have to be specified upfront, along with their characteristics. This is due to a simple reason — using a bottom-up approach yields information about abnormal metrics, however it is very difficult to tell what design defect they actually point towards [37].

As already mentioned, design defects are often defined in a very loose manner. Therefore, in order to automatically detect them, it is necessary to first transform their loose, textual definitions into quantifiable, concrete rules. Operating on the level of source code has the benefit of full availability of technical information. However, design defects are defined on a design level. Thus, to detect design level problems in the source code, it is necessary to first abstract from the concrete implementation [3, 11, 22].

**Does Python Smell Like Java?**

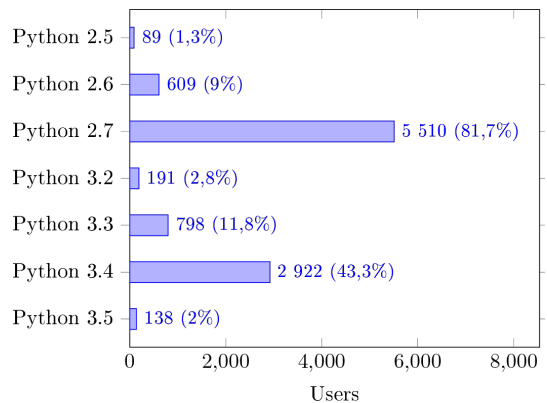

**Figure 1** Usage of Python versions among programmers in 2014 [8].

A central consideration in the design of the Design Defect Detector is which Python version(s) should be supported. Python as a language is still evolving and as of 2017, multiple different versions are being actively used by the developers. An online survey [8] about the year 2014 with 6,746 respondents has shown, that although Python 2.7 and Python 3.4 were the most widely used versions of Python at that time, other versions were still rather common, as can be seen in figure 1.

Some of the available Python versions have major differences between them: for instance, Python 3.x is not backwards compatible with Python 2.x. Distinguishing between the versions is a very time consuming task which seems impossible to automate. Most projects on GitHub, which is the source of our data set, do not state which precise version of Python are they written in. Many projects on PyPi, which we crawl to get the relevant GitHub projects, belong to 5–10 different categories of "Programming Language :: Python :: ...", which are not reliable for at least these four reasons:

1. The categories are added manually by the authors and not checked automatically.
2. A particular version not being listed simply means the authors did not try that version, so the project cannot be used as a negative test case.
3. It is not uncommon for projects to contains several loosely connected scripts, in which case some of them may use a different Python version than others.
4. Many projects strive to support both Python 2 and Python 3. Their codebases are either compatible with both, or moving in that direction.

There are no automatic tools that perform this task, and there cannot be, since some constructs like print('a',) are parsable as either Python 2 or Python 3, but with different semantics. All these issues have forced us to develop an over-approximating parser for a superset covering all currently used versions of Python. The process around this endeavour is explained in section 2.

Within this project, we intend to seek answers to the following research questions:

1. Which of the well known design defects can be detected in Python code?
2. Do Java code and Python code have comparable design defects?

3. What is the density of these design defects in Python code?
4. Do Java code and Python code have comparable design defect density?

These research questions will be answered in section 5.2. For now, we follow up with section 1.3 which explains the line of prior research that this paper strives to continue and contribute to. Section 2 will explain the process of grammar engineering to deliver an over-approximating ANTLR4 grammar capable of parsing millions of lines of code written in Python of different versions. Section 3 contains information about the source code model that is built from the parse trees and forms the foundation for the design defects to be found. The harmful patterns themselves are explained, defined and formalised in section 4 — we mostly borrow the definitions from classic books [7, 18] and construct specifications based on implementations from previously published papers or inspired by them. Evaluating our tool starts in section 5: we explain how the huge test data set of Python modules was harvested and filtered, and present the result of the evaluation in table 5. Traditionally, section 5.3 lists some possible threats to validity, while section 6 concludes the paper by recalling the main contributions, discussing them and sketching possible future work initiatives.

### 1.3 Related Work

The commonly mentioned and researched design defects are code smells [18] and antipatterns [7]. Since their introduction, multitude of different methods for their detection (and sometimes also correction) has been proposed in the literature. They range from manual approaches to semi-automatic or automatic ones.

Mäntylä et al. [36] use one of the manual approaches for detecting bad code smells. They evaluate developer questionnaires about their occurrence and then explore correlation between the smells. A different manual approach was used by Moha et al. [46], who have classified design pattern defects into groups. Their work is based on the definition of design defects being design patterns implemented in a wrong way. They have shown presence of these design defects by having students look for design patterns in code and examining the distortion of these design patterns.

Dhambri et al. [14] propose a semi-automatic method for detecting design flaws. They automatically detect certain symptoms and visualise them, but rely on a human analyst to draw final conclusions.

Ciupke [11] proposes an automatic method which queries a metamodel of the source code for design problems. Guéhéneuc and Albin-Amiot [23] introduce a method for automatically detecting and correcting inter-class design defects. They identify distorted forms of design patterns by using constraint relaxation on a source code meta-model and apply transformation rules based on the relaxed constraints to fix the defects.

An automatic method, which includes visualisation of the results, was proposed by van Emden and Moonen [15]. It is based on a source code model, which stores primitive smell aspects. Alternative approaches for detecting design defects based on source code metrics were proposed by Marinescu [37], Munro [47], and Fontana and Maggioni [16].

The extensive work of Moha et al. lays out an automatic way to detect design defects using so called "rule cards" and a DSL [42] and results in DECOR [40, 41, 43, 44], the state of the art method to automatically generate design defect detection algorithms. Khomh et al. [26] propose an approach which extends DECOR and accommodates for uncertainty by using Bayesian Belief Networks to rank and prioritise classes based on their probability of being part of an antipattern.

Oliveto et al. [48] automatically identify antipatterns through using B-Splines — interpolation curves built using a set of metrics and their values for a given class. Antipatterns are detected based on the class' B-Spline distance from known antipattern classes' B-Splines and known good quality classes' B-Splines.

A logic-based approach was proposed by Stoianov and Sora [57] who define Prolog rules to automatically detect design patterns and antipatterns.

An approach to automatically detect and correct design defect based on genetic programming has been proposed by Kessentini et al. [25] This approach, unlike most others, does not require upfront definition of detection rules.

Most of the methods proposed in the literature, are heavily language dependent. One of the pleasant exceptions is an approach proposed by Llano and Pooley [33], who define UML specifications for antipatterns at a design level and guidelines for manual refactoring of these antipatterns. Another language independent approach was developed by Cortellessa et al. [12]. They formalise performance antipatterns and use the OCL specification language to describe expressions on UML models and query them. Lastly, an approach to automatic refactoring of design defects based on relational algebra is described by Moha et al. [45].

## 2 Parsing Python

The first step in the Design Defect Detector workflow performs parsing the source code of the Python projects. Parsing in a broad sense means recognising structure in textual input [68], so creating a code model that we will explain in the next section, is also a part of the same process, but this section will be about parsing in a narrow sense: creating a hierarchical (tree-like) representation of the flat textual input — in our case, Python programs. Since writing a parser manually is a tedious and a time consuming task [1, 2, 21, 22], an automatic parser generator was used for this purpose. It should be noted that this does not make parsing less challenging, but just automates simpler and more tedious aspects of the work, making the result less error-prone. There are many parser generators available, but we decided to use ANTLR4 [50, 51], as there was an available Python 3.3.5 grammar, written by Bart Kiers [27].

However, this grammar alone was insufficient for the Design Defect Detector, because as explained before, supporting a single version of Python would make the tool applicable only under very specific conditions, which are apriori undetectable. To achieve wider support, we have combined the grammars for different Python versions into a single grammar. Bart Kiers' version was used as a baseline. The remaining resources include the official Python documentation pages [54, 55], which offer grammar specifications for the different Python versions (complete yet not executable),

■ **Table 1** Convergence of Python 2.7 and Python 3.3 grammars, a simple case.

| Python 2.7 specification | Python 2.7 ANTLR grammar | Python 3.3 ANTLR grammar | Converged 2.7 & 3.3 ANTLR grammar |
|---|---|---|---|
| small_stmt | small_stmt | small_stmt | small_stmt |
| : ( expr_stmt | : expr_stmt | : expr_stmt | : expr_stmt |
| \| print_stmt | \| print_stmt |  | \| print_stmt |
| \| del_stmt | \| del_stmt | \| del_stmt | \| del_stmt |
| \| pass_stmt | \| pass_stmt | \| pass_stmt | \| pass_stmt |
| \| flow_stmt | \| flow_stmt | \| flow_stmt | \| flow_stmt |
| \| import_stmt | \| import_stmt | \| import_stmt | \| import_stmt |
| \| global_stmt | \| global_stmt | \| global_stmt | \| global_stmt |
| \| exec_stmt | \| exec_stmt |  | \| exec_stmt |
|  |  | \| nonlocal_stmt | \| nonlocal_stmt |
| \| assert_stmt | \| assert_stmt | \| assert_stmt | \| assert_stmt |
| ) | ; | ; | ; |

and ANTLR3 grammar for Python 2.5 authored by Frank Wierzbicki [64] which was in turn based on the ANTLR2 grammar for Python 2.3.3 by Terence Parr and Loring Craymer [52]. The original grammar was automatically migrated from Python 2 documentation [54], combined with Craymer's lexer and hand tweaked for lookahead correctness by Parr [52].

To create a grammar that covers all versions of Python starting from 2.5, we used a grammar adaptation method known as *grammar programming* [13]. Essentially, this iterative process consists of converging two existing grammars into a single, over-approximating one. The resulting grammar covers a superset of both languages described by the original grammars. The techniques of grammar *adaptation* and grammar *convergence* are highly non-trivial and explained in greater detail by Lämmel [28] and Lämmel and Zaytsev [30] respectively. In short, programming a grammar is a more robust and maintainable way of building a parser and ultimately a compiler, but it does not completely let one abstract away from the problem complexity.

The Kiers' grammar [27] served as a baseline to be extended. To combine the grammars together, we translated the available Python 2.5 grammar [64] from ANTLR3 into ANTLR4 and subsequently merged it with the base grammar of Python 3.3. Afterwards, all the other grammar specifications from the documentation of Python 2 [54] and Python 3 [55] were compared to the base grammar, the necessary parts were translated to ANTLR4 and merged in as well. In case of conflicts we would always as a matter of principle choose the larger alternative or program an over-approximation, since our grammar is meant to be used for analysis and not for correctness verification of source code.

In general, the difficulty of the combination varied. Some parts were very straightforward, such as the definition of small_stmt rule, see table 1. However, some of the rules differed more from each other and the differences did not span across a single rule, but also its subrules and/or rules dependent on it, see table 2.

There is also an issue with reserved keywords: *print* and *exec* are reserved keywords in Python 2.x, but not in Python 3.x, while *nonlocal* is a reserved keyword in Python 3.x

**Does Python Smell Like Java?**

■ **Table 2** Convergence of Python 2.7 and Python 3.3 grammars, a more challenging case.

| Python 2.7 specification | Python 2.7 ANTLR grammar | Python 3.3 ANTLR grammar | Converged 2.7 & 3.3 ANTLR grammar |
|---|---|---|---|
| list_for<br>: 'for' exprlist 'in'<br>testlist_safe<br><br>[list_iter] | list_for<br>: FOR exprlist IN<br>testlist_safe<br><br>list_iter?<br>; | comp_for<br>: FOR exprlist IN<br>or_test<br><br>comp_iter?<br>; | comp_for<br>: FOR exprlist IN<br>test_nocond<br>((',' test_nocond)+ '.'?)?<br>comp_iter?<br>; |
| testlist_safe: old_test<br>[(',' old_test)+ [',']] | testlist_safe: old_test<br>((',' old_test)+ ','?)?<br>; | | |
| list_iter<br>: list_for<br>\| list_if | list_iter<br>: list_for<br>list_if<br>; | comp_iter<br>: comp_for<br>\| comp_if<br>; | comp_iter<br>: comp_for<br>\| comp_if<br>; |
| list_if<br>: 'if' old_test<br>[ list_iter ] | list_if<br>: IF old_test<br>list_iter?<br>; | comp_if<br>: IF test_nocond<br>comp_iter?<br>; | comp_if<br>: IF test_nocond<br>comp_iter?<br>; |
| old_test<br>: or_test<br>\| old_lambdef | old_test<br>: or_test<br>\| old_lambdef<br>; | test_nocond<br>: or_test<br>\| lambdef_nocond<br>; | test_nocond<br>: or_test<br>\| lambdef_nocond<br>; |
| old_lambdef<br>: 'lambda'<br>[ varargslist ]<br>':' old_test | old_lambdef<br>: LAMBDA<br>varargslist?<br>':' old_test<br>; | lambdef_nocond<br>: LAMBDA<br>varargslist?<br>':' test_nocond<br>; | lambdef_nocond<br>: LAMBDA<br>varargslist?<br>':' test_nocond<br>; |

but not in Python 2.x. In addition, Python 3.5 introduced keywords *async* and *await*. To account for this disparity, we added a production rule for identifier, which covered the standard identifiers shared among all versions, but also *print, exec, nonlocal, async* and *await*.

To test the extent of the combined grammar's capabilities, we used a data set of more than 30 million lines of Python code. This data set is further described in section 5. This code was parsed by the ANTLR generated parser. Each file that did not parse correctly, was manually inspected afterwards. During the inspection, the file was categorised as either (1) not conforming to any of the supported grammars or (2) conforming to one or more of the supported grammars. The files which belonged to the second category were collected and combined into a smaller data set, designated for quick testing of the discovered grammar issues and their fixes. All of the discovered issues were fixed. This testing process was used to polish the final grammar, not to guide the entire grammar programming effort which could have been an alternative route take, since it is known to produce good results [4, 62]. Instead, we focused on covering the language features documented in the official language manuals [54, 55] and used the test suites mainly as a validation instrument.

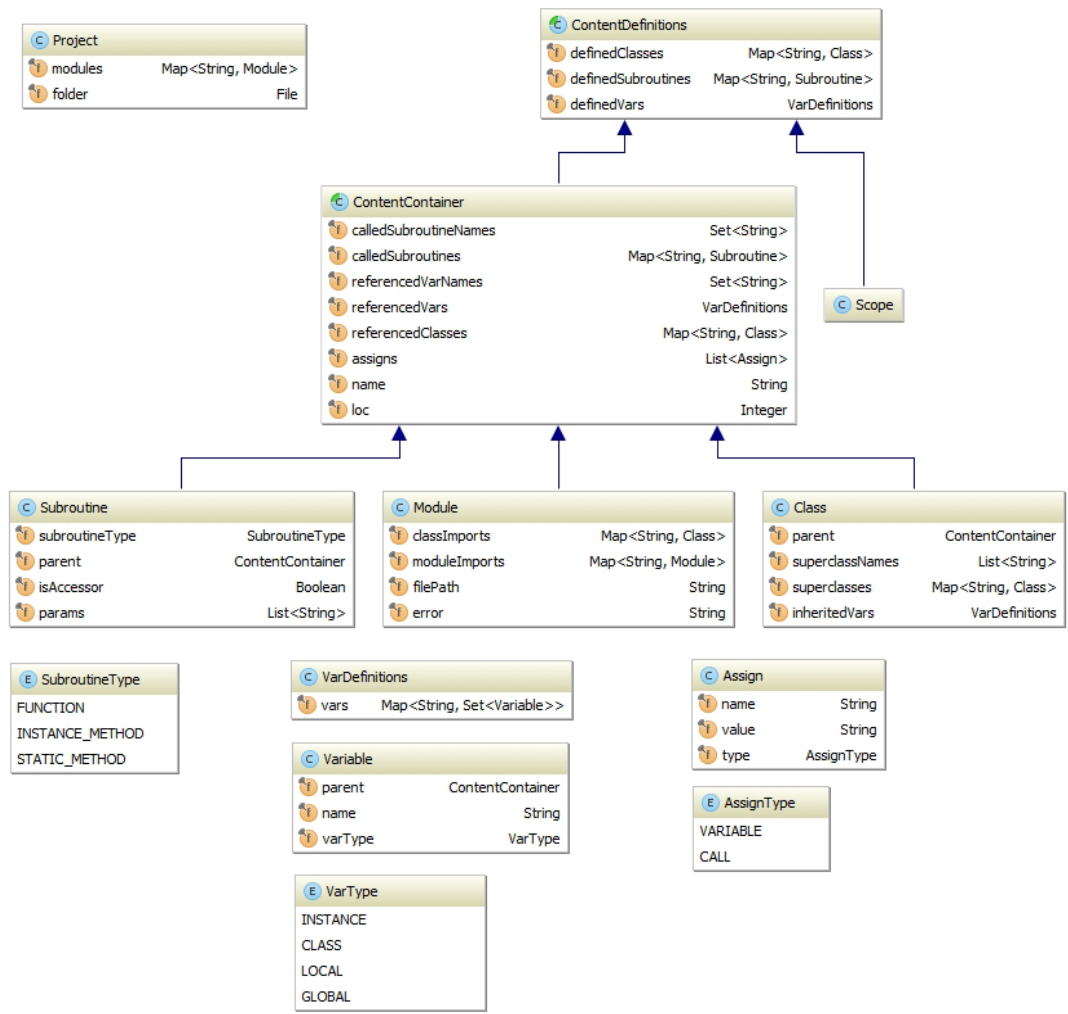

**Figure 2** Source code model structure.

The resulting grammar is a **level 4** grammar in the quality model of Lämmel and Verhoef [29]. This means it can be used to produce a realistic parser, which has been used to parse several million lines of code. To the best of our knowledge, this is the first and only such Python grammar in existence. The converged grammar consists of **143** terminal symbols, **217** nonterminal symbols and **394** production rules (following the standard definitions of TERM, VAR and PROD metrics [9] as calculated by GrammarLab [67]). The grammar and its 24 "development versions" [4], are available publicly at https://github.com/nvavrova/thesis/blob/master/Python.g4.

## 3 Code Model Construction

Building an abstract syntax tree (AST) from the parse tree delivered by ANTLR, is a straightforward process that would not surprise anyone familiar with parsing technology in general and ANTLR4 in particular, and while being a decent piece of engineering everyone is encouraged to inspect, it has a chance of boring the readers beyond acceptable levels. However, the construction of the *code model* goes further than simply abstracting from the concrete code representation. The main purpose of the code model is to simplify querying for the necessary metrics and other characteristics by linking together the different parts (such as classes and variables). Building the code model is therefore based on *fact extraction*, which is explained in detail by Lin and Holt [32].

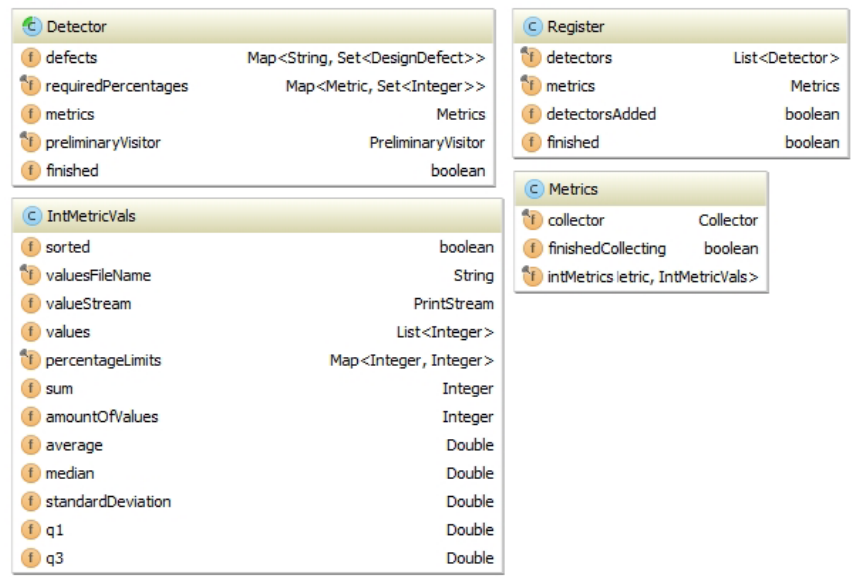

**Figure 3** Simplified overview of the analyser classes.

The code model is a structure inspired by the general structure of any project based on the object-oriented paradigm. At the heart of the code model there are four classes: Project, Module, Class and Subroutine. The code model structure can be visually inspected on figure 2. Its individual parts are also explained below.

**Project** is basically the root node of the code model heart, and at the same time it is the simplest of all the code model parts. Its main purpose is to aggregate the modules based on their location in the folder structure.

**ContentDefinitions** is an abstract class that contains the definitions for different Classes, Subroutines and Variables.

**ContentContainer** is an abstract class which in addition to the definitions also contains the references to different Classes, Subroutines and Variables.

**Module** apart from being an implementation of ContentContainer, also holds the references to all the Classes and Modules that were imported inside of it.

**Class** is another implementation of ContentContainer, and the most heavily used object during the analysis of the code model. A lot of OO design defects are defined on the class level, thus classes play a crucial role in the process of detecting them. In our code model, a class holds references to its parent object (e.g., Module) and to its superclasses.

**Subroutine** is the simplest type of a ContentContainer. The most important task of Subroutines is holding a reference to their parent object (i.e., Class or Module).

The construction of the code model mostly works with the provided AST. It utilises ANTLR's AST visitor and requires two passes. The first pass focuses on the primitive properties, i.e. those that can be directly observed in code. An example of a primitive property is the name of the variable used in a particular class. The second pass obtains the derived properties, i.e. properties that are inferred from other properties. An example of a derived property is a reference to a class that was imported by a specific module.

The first pass constructs a skeleton tree and fills it with simple, observable information. The AST visitor creates the individual classes and adds the important information, such as names of superclasses, etc. The second pass consists mostly of linking the model parts to each other. A separate, smaller AST visitor is used during the second phase. It visits the import statements and based on the existing skeleton, creates references from one class to another. After its completion, it continues to resolve the dependencies such as referencing a variable of a different class. This process is facilitated by creating a Scope in each Module and passing it down to the defined Classes and Subroutines.

After the code model is constructed, the code analyser uses it to detect design defects. This constitutes the final step of the Design Defect Detector. The most important components of the analyser are shown on figure 3 and explained below.

**Metrics** is the class that holds the knowledge about all the metrics that are required for the semantic relative filters or statistical filters (based on percentiles and outliers, as further detailed in section 4). It also supplies the knowledge about the limits for these filters after all the metrics are stored.

**IntMetricVals** performs the heavy lifting in terms of calculating the necessary limits for metric filters. Our implementation only uses integer metrics, such as LOC. However, it could be easily extended to different types.

**Detectors.** Each design defect has its own, separate detector. Detector is the component which decides whether the object (class or subroutine) is a design defect or not. This decision is a two step process, facilitated by the Metrics.

In the first step, the detector checks whether the object possesses properties of the given design defect which are easily observable from the code model (e.g., if the class uses global variables). If yes, it is judged to be a candidate for a design defect and Detector stores the necessary metrics specific to this object for further inspection.

The second step occurs after all the objects of all projects have been through the first step. The Detector then checks all the design defect candidates and compares their metrics against the limits for semantic relative filters or statistical filters, supplied by the Metrics. At this point, the Detector either confirms or rejects the object as a design defect.

**Register** is a simple class that facilitates registering all the Detectors and supplies them with Metrics. It also triggers the metric collection phase and finalisation phase.

| Data Filter | Specification | | Filter Example |
|---|---|---|---|
| Marginal | Semantic | Relative | *TopValues*(10) |
| | | | *BottomValues*(5%) |
| | | Absolute | *HigherThan*(20) |
| | | | *LowerThan*(6) |
| | Statistical | | *BoxPlot* |
| Interval | Composition of two marginal filters, with semantic limit specifiers of opposite polarities | | *Between*(20, 30) := *HigherThan*(20) ∧ *LowerThan*(30) |

■ **Figure 4** Classifications of various types of data filters as defined by Marinescu [37].

Analysis was technically the most challenging part to implement due to the large quantities of source code processed by Design Defect Detector. The analyser had to go through the code model of each individual project and store required information about it. Since the data was way too large to be stored in memory, our implementation extensively uses the file system. The information about all collected metrics and also about potential design defects was stored in files. For this purpose, we have implemented file based mapping collections SetStrMap, SetIntMap, ListMap and PrimitiveIntMap. For running Design Defect Detector on our data set of about 32 million LOC, the necessary temporary data reached about 2,5 GB in size.

## 4 Design Defect Detection

To detect the design defects, we have used mechanisms of filtering and composition of metrics [37] to define the design defect detection rules in a quantifiable manner. Composition is a simple application of logical operations, such as *and*, *or* and *not*. Filtering is based on applying different types of filters, the classification of which can be seen in figure 4. For the Design Defect Detector, we mostly use a relative semantic percentage based filter, e.g. TopValues(10%), and statistical filter, i.e. BoxPlot with 2 types of considered outliers, mild and extreme.

In our definitions of design defects, we use the following functions:

**MildOutlier($M$)** — the metric $M$ is a mild outlier among all values of $M$ measured within a project. Mild outliers are at least 1,5 interquartile ranges away from the median.

**ExtremeOutlier($M$)** — the metric $M$ is an extreme outlier among all measured values of $M$. Extreme outliers are at least 3 interquartile ranges away from the median.

**Top$X$Percent($M$)** — the metric $M$ is in the top $X$ percent of all measured values of $M$.

The design defects detected by Design Defect Detector are listed in table 3.

■ **Table 3** Design defects detected by Design Defect Detector

| Design Defect | See | Type | Definition | Implementation |
|---|---|---|---|---|
| Feature Envy | section 4.1 | Code smell | [18] | [31] |
| Data Class | section 4.2 | Code smell | [18] | |
| Long Method | section 4.3 | Code smell | [18] | [44] |
| Long Parameter List | section 4.4 | Code smell | [18] | [44] |
| Large Class | section 4.5 | Code smell | [18] | [44] |
| God Class | section 4.6 | Antipattern | [7, 53] | [41] |
| Swiss Army Knife | section 4.7 | Antipattern | [7] | [41] |
| Functional Decomposition | section 4.8 | Antipattern | [7] | [41] |
| Spaghetti Code | section 4.9 | Antipattern | [7] | [41] |

## 4.1 Feature Envy

**Description:** Feature Envy is a smell of a method that seems more interested in data of a different class than the one it is in. This often means invoking a large amount of accessor methods.

**Implementation:**

$$(AID > 4) \wedge (Top10\%(AID)) \wedge (ALD > 3) \wedge (NRC < 3)$$

where:
- AID (Access of Import Data) is the amount of referenced variables that do not belong to this class;
- ALD (Access of Local Data) is the amount of referenced variables that do belong to this class;
- NRC (Number of Related Classes) is the amount of different classes referenced in this one.

## 4.2 Data Class

**Description:** Data Class is a smell of a class that only contains data fields and accessor/mutator methods for these fields. Data Class is used solely for the purpose of holding data.

**Implementation:**

$$ExtremeOutlier(AOPuF) \vee ExtremeOutlier(AOA)$$

where:
- AOPuF (Amount of Public Fields) is the number of fields that are made public;
- AOA (Amount Of Accessors) is the number of methods which main purpose is to provide access to the object's fields.

### 4.3 Long Method

**Description:** Long Method is a smell of a method that is too long and should be decomposed into smaller pieces. Fowler et al. already stated that comments (signalling some kind of semantic distance), conditionals or loops often signify a place where the code should be extracted to a separate method.

**Implementation:**
$$ExtremeOutlier(LOC)$$

where LOC (Lines Of Code) is the amount lines of code for the given method, excluding comments and empty lines.

### 4.4 Long Parameter List

**Description:** Long Parameter List is a smell of a method which definition has too many parameters.

**Implementation:**
$$ExtremeOutlier(NOP)$$

where NOP (Number Of Parameters) is the number of arguments the given function takes.

### 4.5 Large Class

**Description:** Large Class is a smell of a single class trying to do too much. The instance variables of this class are used in a lot of places, possibly as part of duplicated code.

**Implementation:**
$$ExtremeOutlier(NMD + NAD)$$

where
- NMD (Number of Methods Defined) is the number of concrete methods in a class;
- NAD (Number of Attributes Defined) is the number of fields and properties in a class.

### 4.6 God Class

**Description:** God Class, also known as Blob, is an antipattern that consists of one complex class, possibly surrounded by multiple Data Classes. The complex class monopolises all the processing, while the only responsibility of the Data Classes is to encapsulate data.

**Implementation:**
$$IsController \wedge MildOutlier(LOC) \wedge MildOutlier(LCOM) \wedge (RDC > 2)$$
where

- *IsController* = (*HasControllerName* ∨ *HasControllerMethods*);
- *HasControllerName* holds if the class' name contains words such as Manage, Process, Control, etc.;
- *HasControllerMethods* holds if the class contains methods that have a controller name;
- LOC (Lines Of Code) is the number of lines of code for the given class, excluding comments and empty lines;
- LCOM (Lack of Cohesion of Methods) is the classic OO metric as defined by Chidamber and Kemerer [10];
- RDC (Related Data Classes) is the number of classes that have *Top15%*(*AOA*);
- AOA (Amount Of Accessors) is the number of methods which main purpose is to provide access to the object's fields.

### 4.7 Swiss Army Knife

**Description:**  Swiss Army Knife is an antipattern for a class with too many responsibilities. This can easily be observed by not just having a large number of methods, but in particularly implementing too many interfaces and/or using multiple inheritance.

**Implementation:**
$$ExtremeOutlier(SUP)$$
where SUP is the amount of superclasses of the given class.

### 4.8 Functional Decomposition

**Description:**  Functional Decomposition is an antipattern for a class that does not leverage object-oriented principles such as inheritance and polymorphism. It usually has a single action as a function and all its attributes are private and used only inside the class. It often has a function-like name (e.g., CalculateInterest).

**Implementation:**
$$HasProceduralName \wedge (SUP = 0) \wedge (RCOMPF > 2)$$
where

- *HasProceduralName* holds if the class' name contains words such as Make, Create, Exec, etc.;
- SUP is the amount of superclasses of the given class;
- RCOMPF is the amount of Related Classes with One Method and a lot of Private Fields, i.e. classes that only have one method and have *MildOutlier*(*AOPrF*);
- AOPrF (Amount of Private Fields) is the number of fields that are made private.

**Does Python Smell Like Java?**

### 4.9 Spaghetti Code

**Description:** Spaghetti Code is an antipattern for a piece of code lacking any structure. This is often detectable by the use of global variables instead of parameters for methods. Spaghetti Code does not make use of basic OO concepts such as inheritance or polymorphism.

**Implementation:**

$$HasProceduralName \wedge (SUP = 0) \wedge UsesGlobals \wedge HasLongMethod \wedge (Top15\%(MNP))$$

where

- *HasProceduralName* holds if the class' name contains words such as Make, Create, Exec, etc.;
- SUP is the amount of superclasses of the given class;
- *UsesGlobals* holds if at least one of the variables referenced by this class is a global variable;
- *HasLongMethod* holds if at least one of the methods of this class has $Top15\%(LOC)$;
- LOC (Lines Of Code) is the number of lines of code for the given class, excluding comments and empty lines.
- MNP (Methods with No Parameters) is the number of methods with $NOP = 0$.

## 5 Evaluation

To obtain the data set used in our experiment, we have used PyPI in combination with GitHub and GitHub API. PyPI, or Python Package Index, is the official repository of open-source, third party software for Python. It can be found at https://pypi.python.org/pypi. GitHub is a popular web-based repository hosting service based on a version control system called Git. Git is explained in detail by Loeliger [34], and GitHub API can be found at https://developer.github.com/v3/.

Initially, we automatically collected a comprehensive list of all GitHub links belonging to any PyPI package submitted under the category Python 2.5 and above. We have obtained 17,568 unique GitHub repository links.

To ensure that the data set is suitable for the purposes of this research, this list was filtered, ensuring that each repository:

- is available and cloneable: public, not deleted from GitHub, not renamed, has no overly long names that stress the OS limits and no files like aux.py that prevent it to be cloned under Windows.
- is not a fork. As one of our objectives is to find out the density of design defects, it would be detrimental to include multiple forks of the same project into our research. To prevent this, we queried the GitHub API for information about the collected GitHub repositories. Any fork repository was removed from the result set and instead replaced with its parent (recursively).

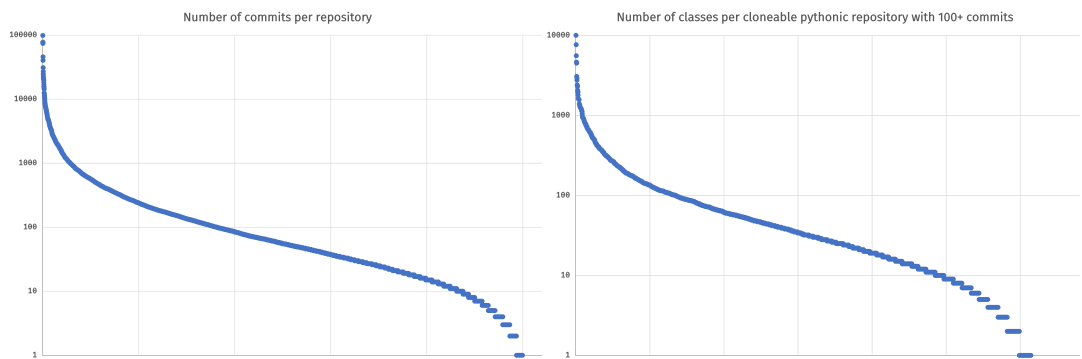

**Figure 5** On the *left*: number of commits per repository found at PyPI. On the *right*: number of classes per cloneable repository with 100+ commits and mostly Python code. The scale on the Y axes on both plots is logarithmic.

- is a Python repository in a sense that either 40% or more of the code in it is in Python, or the usage of each of the non-Python languages individually is lower than that of Python. For example, @numpy/numpy consists of 52.7% C code, but 46% Python code, which we consider enough for it to be a legitimate Python repository. However, @aosabook/500lines contains 28.9% Python, but 49.4% JavaScript, so we filter it out.
- has been in active development at least for some time, having **100** or more commits. A plot of the number of commits per all repositories after applying filtering rules above, can be seen in figure 5 (left).
- is sufficiently large, having **20** or more classes. This requirement was applied because a large portion of design defects which we are trying to detect are related to object-oriented programming. If the project is functional or scripted for most part, it is irrelevant to our research. Figure 5 (right) plots the number of classes per each of the repositories not filtered out so far.
- has a parse ratio upward of 99%. If the project does not parse correctly, it is not possible to infer the relationships between different classes, methods and variables and therefore it is also impossible to perform the design defect detection.

After filtering out 3,307 unavailable, fork or non-pythonic projects, as well as 10,140 underdeveloped projects, we ended up with 4,121 projects on our hands. Then, we have excluded "hidden" directories with names that start with a dot, since some projects like @jacebrowning/template-python-demo contained a .venv directory that included 87 other packages, promptly bringing its class count from actual 2 to a staggering 5,594 (which would have put the project as the third largest OOP-wise after @python/cpython and @django/django). The selected projects contained in total 238,861 modules, out of which 238,503 were parsed properly (99.85%). Manual inspection of remaining unparsable modules revealed that they are indeed incorrect and not accepted by the Python interpreter. Removing those broken modules from our set left us with 238,503 modules containing 488,008 classes and 32,058,823 LOC.

Some statistics about the resulting data set can be found in figure 6. The biggest projects, either by the number of commits or by the amount of classes, were chosen

**Does Python Smell Like Java?**

■ **Table 4** Biggest projects from our dataset: merged top 10 by the number of commits, top 10 by the number of classes in production code and top 10 by the number of classes in test code.

| Repository | Commits | Classes | Test Classes |
| --- | ---: | ---: | ---: |
| @python/cpython | **98,751** | **2,830** | 7,241 |
| @edx/edx-platform | **40,292** | **1,969** | 2,537 |
| @sympy/sympy | **26,991** | 1,260 | 126 |
| @gem/oq-engine | **24,720** | 145 | 108 |
| @django/django | **24,045** | 1,605 | 6,082 |
| @ipython/ipython | **22,450** | 307 | 190 |
| @scikit-learn/scikit-learn | **21,700** | 519 | 111 |
| @matplotlib/matplotlib | **21,112** | 1,054 | 150 |
| @Abjad/abjad | **20,129** | 541 | 65 |
| @nijel/weblate | **19,844** | 437 | 219 |
| @scipy/scipy | 16,935 | 693 | **921** |
| @numpy/numpy | 15,630 | 341 | **900** |
| @statsmodels/statsmodels | 8,916 | 600 | **977** |
| @ManageIQ/integration-tests | 8,113 | 0 | **1,274** |
| @tomchristie/django-rest-framework | 7,093 | 196 | **989** |
| @Tapyr/tapyr | 5,751 | **2,937** | 139 |
| @emilkarlen/exactly | 2,850 | 498 | **1,101** |
| @wxWidgets/Phoenix | 2,828 | **1,843** | 517 |
| @MongoEngine/mongoengine | 2,779 | 103 | **1,111** |
| @jleclanche/fireplace | 2,552 | 1,590 | 1 |
| @Azure/azure-sdk-for-python | 1,867 | **2,263** | 139 |
| @google/grr | 805 | **2,869** | 88 |
| @pydcs/dcs | 611 | **2,767** | 7 |
| @rwl/PyCIM | 153 | **2,309** | 5 |

to be listed in table 4. To separate test code from production code, we relied on file names and directory structure: if the filename or the path to it contained "test" or "Test", all the classes found in that file were classified as test classes.

We ran the Design Defect Detector [61] on the data set of these 32 million lines of Python code. The amount of detected design defects and their density per design defect type is shown in table 5. One can see that we have found:

- No occurrences at all of Functional Decomposition and barely any of God Class and Spaghetti Code;
- Significantly lower density of Long Parameter Lists in Python than in Java;
- Significantly higher density of Long Methods in Python than in Java;
- Relatively comparable densities for Large Classes and Swiss Army Knives.

The measured average density of a design defect per 10,000 lines of code for Python and Java were 6.07 and 8.37 respectively.

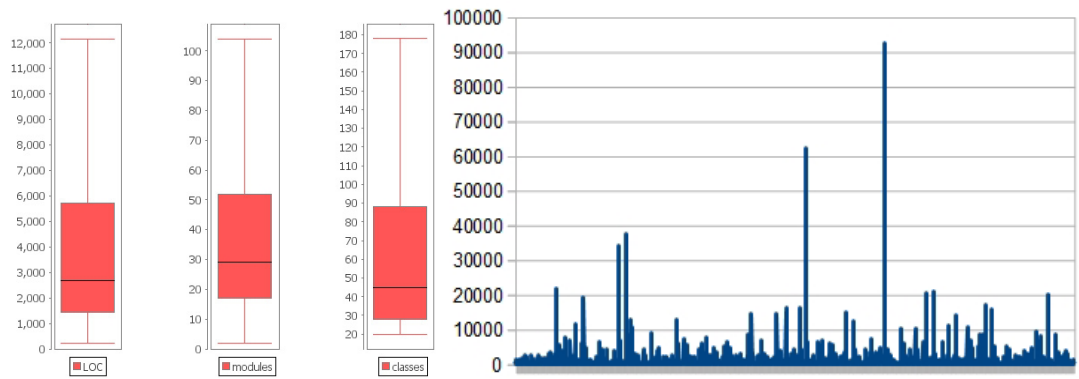

**Figure 6** On the *left*, distribution per project of LOC, number of modules and number of classes. On the *right*, commit counts per project.

## 5.1 Discussion

Python is generally considered much more concise than Java, meaning that developers can accomplish the same task with less lines of code. Our results show that in most cases, the design defect density per 10,000 lines of code in Python is lower than that in Java. This result may suggest that Python is superior to Java in the aspect of design defects. However, there are arguments which undermine this theory, such as that the concrete definitions which work for Java source code are insufficient in the context of Python and they need to be tailored for this specific context.

Another interesting fact is that in our experiment, detection rate of code smells in Python was higher than that of antipatterns (7.80 per smell on average vs. 2.34 per antipattern). On one hand, this could mean that while Python code contains low level issues just like Java code, it has less architectural problems compared to Java code. On the other hand, the numbers are so low for some of the antipatterns that it might suggest that Python code and Java code just have *different* architectural level antipatterns. We will go deeper into details of threats to validity and future work in section 5.3 and section 6, respectively.

The relative abundance of Long Methods that we have found in Python code, may be attributed to at least two different reasons. The first possible explanation is a common scenario of object-oriented Python code being evolved from a functional or imperative script that was getting out of hand. This way, conceptual units of programmer's thinking remain intact and the blob gets split only in places that seem really necessary. The existing literature also agrees that in many cases long methods are faster to write and easier to maintain [39]. Java programmers have a different workflow and tend to start with a basic OO design — for example, a project can be started by creating a significant number of files with classes that do not do anything but sketch the desired hierarchy. The second explanation is purely statistical: it is not uncommon for a Python class to have a lot of one-liner methods; which automatically make any method longer than a screen, an extreme outlier.

Similarly, the results in detecting Functional Decomposition can also be explained by language differences. Java did not have support for anonymous functions until

**Does Python Smell Like Java?**

■ **Table 5** Evaluation and replication: for our tool and for DECOR side by side we show the amount of instances found, the LOC count of the experiment and the density (average number of defects per 10,000 LOC). Data from the smaller of the DECOR experiments comes from [44], data from the larger one from [41].

| | DDD [61] | | | DECOR [41, 44] | | |
|---|---|---|---|---|---|---|
| Design Defect | Found | LOC | Density | Found | LOC | Density |
| Feature Envy | 5,103 | 32,058,823 | 1.59 | | | |
| Data Class | 13,537 | 32,058,823 | 4.22 | | | |
| Long Method | 79,367 | 32,058,823 | **24.76** | 22 | 21,267 | **10.34** |
| Long Parameter List | 10,429 | 32,058,823 | **3.25** | 43 | 21,267 | **20.22** |
| Large Class | 16,576 | 32,058,823 | **5.17** | 13 | 21,267 | **6.11** |
| God Class | 24 | 32,058,823 | **0.01** | 150 | 516,092 | **2.91** |
| Swiss Army Knife | 30,011 | 32,058,823 | **9.36** | 441 | 516,092 | **8.54** |
| Functional Decomposition | 0 | 32,058,823 | **0.00** | 179 | 516,092 | **3.47** |
| Spaghetti Code | 1 | 32,058,823 | **0.00** | 363 | 516,092 | **7.03** |

relatively recently, and even then it is verbose and in some cases found too cumbersome by programmers. Python had lambda expressions since 1.0, and they were working smoothly until Python 2.7. Lambda expressions remained in Python 3.x after a huge discussion about removing them altogether [56], and even though they were made less comfortable for functional programmers with LISP background, they have never gotten bad enough to define single-method classes instead. So, our hypothesis is that Functional Composition as an antipattern is irrelevant for a language with powerful lambdas and comprehensions like Java 8+ or Python of any version.

Spaghetti Code in our results is barely detectable, which does not necessarily mean that all Python code is nicely structured. On the contrary, it is most probably due to the combination of how the antipattern is defined ([7]'s definition is quite controversial and hardly complete), and how we have collected (and filtered) our dataset. Arguably, Python programmers have much more ways to create messy code than simply using globals and neglecting polymorphism: illogical mixing of paradigms, inconsistent string formatting (e.g., %, {} and f" within one module), excessive playing with underscore methods (__call__, __getitem__ and others) and so on.

Low density of God Classes can be easily explained by the multi-paradigmatic nature of Python: if a developer would like to make a Blob, they often would not bother creating a class for it, it can stay a module with loosely connected functions, or even a script.

### 5.2 Answers to Research Questions

To answer the project's research questions from section 1.2:

1. **Which of the well known design defects can be detected in Python code?**
   We can conclude that the following design defects are detectable: Feature Envy, Data Class, Long Method, Long Parameter List, Large Class, God Class, Spaghetti

Code and Swiss Army Knife. Functional Decomposition seem to not be an issue for Python.

2. **Do Java code and Python code have comparable design defects?**
   Since we have been able to detect the majority of the design defects that have previously been detected by others in Java source code, we conclude that similar design defects can be detected in Java and Python source code.
3. **What is the density of these design defects in Python code?**
   On average, we have detected the density of 6,07 design defects per 10,000 lines of code. The individual measurements per design defect are listed in table 5.
4. **Do Java code and Python code have comparable design defect density?**
   The measured average density of design defects per 10,000 lines of code for Python and Java were 6,07 and 8,37 respectively. The individual measurements per design defect are listed in table 5. This shows that the design defect densities are comparable and according to our results, the density is usually slightly lower in Python code, unless the point of interest is Long Methods.

### 5.3 Threats to Validity

A common threat to validity of all design defect detection studies is introduced by the loose, textual definitions of design defects. Due to their openness to interpretation, personal judgement impacts the selection of suitable metric combinations and their thresholds. This can only be addressed by formal modelling, and we tried to be as open with the formulae we reuse, as possible, enabling future researchers and replicators to take our definitions exactly for what they are.

Another threat is the data set we have used for evaluation. It only consists of public GitHub repositories listed on PyPI. Although we have examined 4,121 repositories, which is quite a lot, we cannot guarantee them to be a representative sample. For instance, there could be fundamental differences between corporate Python code and the open source code available on GitHub, in terms of duplicated code (clones), self-admitted technical debt (to-dos), ownership, etc. One could also argue that GitHub attracts certain types of developers and projects, so future work must include a broader replication with projects from BitBucket, SourceForge, CodePlex, GitLab, RosettaCode, LaunchPad, etc., some of which are relatively small but Python-biased.

The study is also limited by the amount of design defects we have studied. It would require a more extensive study to examine the density of Python design defects in general and make a comparison to Java or any other language. We have paved the way to that research by providing the hardest component to it — the Python grammar, but going there would imply another, albeit easier, project of comparable effort.

Lastly, we have not been able to make any statement about precision or recall of Design Defect Detector. This means that the comparison to results in Java might be skewed one way or the other. If the Design Defect Detector shows a significant number of false negatives, the Python design defect density could be higher than that of Java. On the other hand, if the Design Defect Detector shows a lot of false positives, the design defect density of Python would be lower than what we have measured. This issue would also mean that the definitions successfully used for detecting design defects in Java code first need to be adjusted for our context to work well.

## 6   Conclusion

We have developed the first level 4 [29] grammar of Python, which is capable of parsing both Python 2.x and Python 3.x code. It is a result of a major grammar programming effort, converging previously existing grammars (made for different dialects and written in different notations) [27, 52, 64] and language documentation [54, 55]. To improve and validate the quality of the grammar, we have collected a big test input data set by leveraging 4,121 GitHub repositories for a total of 32,058,823 lines of code in various versions of Python. This is the biggest test data set any Python grammar published up to this day has ever seen. The fact that our grammar was capable of parsing these, provides evidence for its *analytic* (recognising) power, but would have not been enough to claim its correctness.

To provide further evidence of the usefulness of the grammar, we have developed a tool called the Design Defect Detector [61] which parses a Python module, creates a code model of it and reports on the presence of various design defects found there. Density of the found design defects varied per type and was between 0.0 and 24.76 design defects per 10,000 lines of code, with the average density being 6.07. The most commonly found design defects were Long Method (79,367 found instances) and Swiss Army Knife (30,011 found instances). The least commonly found design defects were Spaghetti Code (1 found instance) and God Class (24 found instances). No instances of Functional Decomposition were found. Overall, the measured density was higher for code smells than for antipatterns.

We have aligned our study to serve as a partial replication of DECOR [41, 44], a state of the art tool built for detection of design defects in Java source code. We have successfully compared the detected design defect density in Python to results that DECOR yielded for Java. Generally, the density we have measured in Python (average 6.07 design defects per 10,000 lines of code) was slightly lower than the density measured in Java (average 8.37 design defects per 10,000 lines of code).

Our research has reaffirmed that the context of a programming language is one of the essential factors in software engineering in general and design defect detection in particular. Because of this, the existing research is still rather incomplete and requires studies similar to ours to be conducted on different programming languages in the future.

The possibility of design defects specific to Python should also be examined in future studies. Some of our current research activities involve detecting how "pythonic" a given piece of code is and investigating what are the design defects specific to Python that do not often occur in code written in other programming languages.

Another topic worth exploring in the future would be the relation between the detected design defects in Python and the actual bugginess of the software, similarly to what Taba et al. did for Java [58].

**About the authors**

**Nicole Vavrová** is a senior software engineer with a university degree (Master of Science in Software Engineering, Universiteit van Amsterdam, 2015) and several years of industrial experience in developing web applications from various domains, both solo and as a team player. Contact her at vavrova.n@gmail.com.

**Vadim Zaytsev** is the chief science officer at one of the largest independent compiler companies in the world. His previous work experience includes being a lecturer and a researcher at various academic institutions of the Netherlands and Germany, as well as an active contributor to open source/data projects and communities. Contact him at vadim@grammarware.net.